\documentstyle[aps,pre,multicol,epsf]{revtex}

\begin{document}

\draft

\title{Interface dynamics at the depinning transition}

\author{Juan M. L\'{o}pez$^{1}$ and Miguel A. Rodr\'\i guez$^2$}

\address{$^1$ Department of Mathematics, Imperial College 
180 Queen's Gate, London SW7 2BZ,
United Kingdom
}
\address{
$^2$ Instituto de F\'\i sica de Cantabria (CSIC-UC),
Avenida de Los Castros, E-39005 Santander, Spain.}

\maketitle

\begin{abstract}
We study the local scaling properties of driven interfaces in disordered 
media modeled by the Edwards-Wilkinson equation with quenched noise.  
We find that, due to the super-rough character of the interface 
close to the depinning transition,
the local width exhibits an anomalous temporal scaling. It does not 
saturate at a characteristic time $t_s(l) \sim l^z$ as
expected, but 
suffers an anomalous temporal crossover to a different time
regime $t^{\beta_*}$, where $\beta_* \simeq 0.21 $. 
This is associated with the existence of a local roughness exponent 
$\alpha_{loc} \simeq 1$ that differs from the global one $\alpha \simeq 1.2$.
The relevance of the typical size of pinned sites regions near the critical
point is investigated and the definition of the critical depinning is dicussed. 

\end{abstract}
\pacs{47.55.Mh, 5.40.+j, 68.35.Fx, 05.70.Ln}

\begin{multicols}{2}
\narrowtext

\section{introduction}
Rough surfaces and interfaces appear in many situations
in nature. Deposition, erosion, fluid-fluid
displacement in porous media or fire front motion, are important examples
\cite{general,review} in which
an interface kinetically becomes rough.
A rough interface in dimension $d+1$ (d is the substrate dimension),
which is described by its height $h({\bf x},t)$
at position ${\bf x}$ and time $t$,
has been treated as a self-affine fractal.
In the case of interfaces driven by spatio-temporal noise
the temporal evolution of $h({\bf x},t)$
is given by the stochastic equations of Edwards-Wilkinson (EW) \cite{ew}
and Kardar-Parisi-Zhang \cite{kpz}, which have been widely studied.
The effect of quenched disorder ({\it i.e.} a noise frozen in time)
on the interface dynamics is a more difficult problem and up to now
there is not a complete and consistent picture.
An important feature of the quenched disorder is that the
interface can be dominated by pinning forces,
which can slow down the motion of the interface in large regions.
Pinning phenomena have a considerable importance in problems like the 
immiscible displacement of fluids in porous media or the motion of domain
walls in magnetic systems.   

During the last years much effort has been paid to
this subject, but theory, simulations and experiments are not yet in good
agreement (see \cite{review} for recent reviews).
In this paper we wish to show that the reason for this disagreement is
certainly twofold.
On the one hand,
growth processes do not always generate
truly self-affine interfaces \cite{lack,das-sarma}
and different scaling behaviours may appear at short and large length scales.
On the other hand, the quenched disorder introduces
a new length scale, related to the size of pinned regions that becomes
relevant and, as we will see later, changes the usual scaling laws.

The simplest stochastic differential equation that describes the evolution
of the interface height in the presence of quenched randomness
is the so-called quenched Edwards-Wilkinson (QEW) equation:
\begin{equation}
\label{QEW}
{\partial h \over \partial t} = D {\bf \nabla}^2 h + F + \eta({\bf x},h),
\end{equation}
which is similar to the EW equation
but with a quenched disorder term, $\eta({\bf x},h)$.
The external driving force
$F$ controls completely the dynamics of the front. If $F$ is larger than
a critical force $F_c$, the interface moves with a finite velocity. However,
the interface remains pinned by the disorder for $F<F_c$. The critical
point $F=F_c$ is known as depinning transition. Above the
depinning transition, $F \ge F_{c} $, a characteristic length,
$\xi$, appears.
$\xi$ represents the typical size of the pinned regions
of the interface and close to the transition
scales as $\xi \sim (F-F_{c})^{-\nu}$.
As we will discuss later, $\xi$ is very important
to understand the scaling of the interface near the depinning transition.

The {\em global} interface width, which characterizes the scaling of the
advancing front, is defined by
\begin{equation}
\sigma (L,t) =  \left\{ \langle (h({\bf x},t) -
  \langle h({\bf x},t) \rangle )^2 \rangle^{1/2} \right \},
\end{equation}
where brackets are averages over the whole system 
and $\{$...$\}$ over realizations of the disorder. 
In the usual case,
when the only relevant length is the system size,
the width scales with time for the early times regime and
saturates at long times:
\begin{equation}
\sigma (L,t) \sim
\left\{ \begin{array}{l}
     t^{\beta} \:\:\:\: {\rm if} \:\:\:\:  t \ll t_s\\
     L^{\alpha}\:\:\:\: {\rm if} \:\:\:\:  t \gg t_s,
     \end{array}
\right.
\label{scaling}
\end{equation}
where $\alpha$ is the roughness exponent and $\beta$ is the time exponent.
This scaling picture is known as Family-Vicsek scaling ansatz
\cite{general,review}.
The saturation time $t_s$ depends on the system size $L$ because it is
the time in which the horizontal correlation length $\l_c(t) \sim t^{1/z}$
reaches the system size. The exponent $z$ is called
dynamic exponent and the scaling relation $\alpha = z \beta$ is fulfilled.

An alternative way of determining the roughness exponent $\alpha$ is
frecuently used in simulations because of the long wait needed to get
saturation in large systems, in which simulations are done.
The width is calculated over a window
of length $l \ll L$ and averaged over many pieces with the same length
along the interface. In other words, we can study the scaling of the
{\em local} width
\begin{equation}
\sigma (l,t) = \left \{ \langle (h({\bf x},t) -
  \langle h({\bf x},t) \rangle_l )^2 \rangle_l^{1/2} \right \}.
\end{equation}
The local width scales in a similar manner as the global width
and has a short time regime $\sigma(l,t) \sim t^{\beta}$.
For small length scales (or equivalently at long times),
$l \ll \l_c(t)$, the local width is believed to be independent of time and
to scale as
\begin{equation}
\label{local-rough}
\sigma(l,t) \sim l^{\alpha_{loc}}.
\end{equation}
This method is very useful when one has no possibility of
producing interfaces in different system sizes as occurs in real
experiments. Most theoretical growth models generate {\em self-affine} 
interfaces and $\alpha=\alpha_{loc}$ \cite{general,review}.

Much work has been carried out to obtain the critical exponents
at the depinning transition. The roughness exponent has been
the most controversial.
Experiments performed in $d=1$ \cite{exper}, which are believed
to be described by Eq.(\ref{QEW}), gave roughness exponents
around 0.75-0.9 in disagreement with a
renormalization group prediction $\alpha = 1$ \cite{RG}. A scaling theory
\cite{anomalo} gave an effective roughness exponent 0.8 in
good agreement with the model reported in Ref.\cite{europhys}.
A number of numerical models for interface pinning by quenched
disorder have been investigated. Roux and Hansen \cite{roux}
found two different
exponents: $\alpha_{loc} = 0.86 \pm 0.03$ when the local width was used,
and $\alpha \simeq 1.16-1.20$
looking at the scaling of the global width as a function
of the system size. Results reported by Jensen \cite{jensen},
$\alpha_{loc} \simeq 0.9$ and $\alpha \simeq 1.15$ are similar.
Makse and Amaral \cite{makse} obtained $\alpha_{loc} = 0.92 \pm 0.04$
or $\alpha \simeq 1.23 \pm 0.04$ depending on the method that they used.
Also other numerical determinations of $\alpha_{loc}$ gave values around 0.8
\cite{nolle,kessler,martys,galluccio,csahok}
and $\alpha \simeq 1.2$ \cite{leschhorn,galluccio}.

It has been demonstrated that the reason for the
existence of two different roughness exponents
is the {\em super-rough} character of the interface close to the depinning
transition. In Ref.\cite{leschhorn} was already shown that the local width is 
bounded by $\sigma(l,t) \le l$ and for that, it is just {\em technically}
imposible to measure a roughness exponent larger than one using the local
width. On the contrary the global width does not have this restriction
so that it gives the {\em true} roughness exponent.
However, we would like to note here that there also exist other effects 
\cite{lack,das-sarma} that may lead to the same 
anomalous scaling in surfaces even with $\alpha < 1$. 
Our aim in this paper is to show that the local width
suffers a temporal crossover, instead of saturating, associated with 
a different
interface scaling at short, $l \ll L$, and long, $l \sim L$, length scales.
The scaling properties of the interface cannot be correctly described
by the usual dynamic scaling hypothesis. We will see that a super-rough 
dynamic scaling gives the correct scaling of the front.

\section{numerical results}
We begin by performing a numerical integration of the QEW equation
in $d=1$ to study the scaling behaviour of the local and global widths.
To start we have to discretize the QEW equation in the horizontal direction
\begin{eqnarray}
h_i(t+\Delta) & = & 
h_i(t) + \Delta \left(D (h_{i+1}(t) + h_{i-1}(t) - 2h_i(t)) \right. \\ 
\nonumber
& + & \Delta \left( F + \eta(i,[h_i(t)]) \right),
\label{QEWdisc}
\end{eqnarray}
where the index $i=1, \cdots, L$, being $L$ the size of the system.
$[h_i(t)]$ represents the integer part of $h_i(t)$.
$h_0(t)=h_1(t)$ and $h_{L+1}(t)=h_L(t)$ are the boundary conditions.
This is the usual discretization scheme used by many authors
\cite{kessler,jensen,csahok}.
The random field $\eta(i,j)$ is Gaussian and its correlation
$\langle \eta(i,j) \eta(i',j') \rangle =
g \delta_{i,i'} \delta_{j,j'}$. Hence the noise is
uncorrelated from a cell to another one and stands
for the time of a run.
The time step $\Delta$ 
(we used $\Delta = 0.01$ in all our simulations)
and the size $L$ have to be chosen in
such a way that the QEW equation is obtained
in the continuous limit ($\Delta \to 0$, $L \to \infty$). So, the following
conditions must be verified: $\Delta D \ll 1$, $\Delta F \ll 1$,
and $\Delta g \ll 1$.
One has to check that changing these parameters does not affect
the final numerical result.
Following Kessler {\it et. al.} \cite{kessler}
we distributed disorder sites randomly
with a prefixed density $p$ over the network. So, the disorder is nonzero
with probability $p$ and on these sites disorder is distributed according
to a Gaussian distribution. The scaling exponents do not depend on
the actual
value of $p$ but the crossover effects will be clearer for small values
of $p$. As we were looking for crossover regimes, we used $p=0.1$ in all
our numerical simulations.
We worked with networks of size $L=1000$, and
the results were averaged over 15-50 realizations of disorder.
As expected we found a critical driving force $F_c$ that separates the
pinning and depinning phases. The value of $F_c$ depends on the actual
values of the parameters of simulation ($p$, $D$, $L$). For
$p=0.1$, $\Delta = 0.01$, $D=5$, and $L=1000$ we estimated $F_c \simeq 0.13$.

\subsection{Scaling far from the depinning transition}
Our numerical results indicate
that the global width has two different time regimes.
Close to the
transition, $F \simeq F_c$, the width behaves as $\sigma(L,t) \sim t^\beta$
, where $\beta \simeq 3/4$ is the time exponent near the
critical point. Far from the pinning transition, $F \gg F_c$,
$\sigma(L,t) \sim t^{\beta_{EW}}$ and the time exponent
$\beta_{EW} \simeq 1/4$ corresponds to the time exponent of the
EW equation.
As we will see later, the values of the remaining
exponents $\alpha$ and $z$ indicate that the interface dynamics clearly
belong to the EW universality class in the limit of large
driving forces.

In general, the temporal behaviour of
the global width
is well understood near and far from the depinning transition and it
has been analyzed by many authors \cite{europhys,csahok,makse}.
On the contrary, we focus here on the
scaling behaviour of the local width.
Firstly,
as far as time evolution is concerned, in the
strong pushing regime the local width scales with time for early times
and saturates at time $t_s(l)$ as corresponds to the
Family-Vicsek scaling picture (see Fig.1).
The value of the width at
saturation can be used to determine the roughness exponent. We found that
the saturated local width, $\sigma_{sat}(l)$, scales as
$\sigma_{sat}(l) \sim l^{0.5}$ (see Fig.1 inset).
Thus, for large driving forces the exponents
($\beta \simeq 1/4$ and $\alpha_{loc} \simeq 1/2$)
of the EW universality class are obtained, as
expected \cite{europhys,anomalo,makse}.

\subsection{Local scaling at the depinning transition}
At the critical value, $F=F_c$,
the complete scaling description of the local width
is much less trivial than in the strong pushing limit discussed above.
In Fig.2 we display the time evolution of the local width $\sigma(l,t)$ for
different window sizes $l$. After an early time regime, the local
width crosses over to $t^{\beta_*}$, where the asymptotic time exponent is
$\beta_* \simeq 0.25$. This scaling region becomes shorter as the length
scale is larger, $l \to L$. Moreover, there is no saturation of local
widths at times $t_s(l) \sim l^z$, contrary to what is generally thought
saturation of the local width occurs only when the whole system
saturates ({\it i.e.} $t_s(l) \sim L^z$ for any $l$).
As we did in Fig.1 for the strong pushing limit,
we can use the saturated values of the local width in Fig.2
to determine the local roughness exponent (see Fig.2 inset),
$\sigma_{sat}(l) \sim l^{0.92}$.
From Fig.2 it is clear that the dynamic
behaviour of the interface near the depinning transition
cannot be completely described with a Family-Vicsek scaling ansatz.
According to this scaling picture the local width $\sigma(l,t)$ should
saturate when the horizontal correlation length reaches the size of the
sample, $\l_c(t_s) \sim l$, as occurs in the strong pushing
limit (see Fig.1). However, the local width displayed in Fig.2 scales with
time for length scales $l \ll \l_c(t)$
and a new time exponent $\beta_*$ has to be introduced.

This anomalous time regime of $\sigma(l,t)$ has been observed in other
growth processes and termed the problem of {\em anomalous} kinetic roughening 
in the literature \cite{lack,das-sarma}.
In Ref.\cite{lack} we have shown that for a broad class of
growth processes the scaling for $l \ll \l_c(t)$ is actually given by
\begin{equation}
\label{anomal-scal}
\sigma (l,t) \sim t^{\beta - {\alpha_{loc}/z}} \; l^{\alpha_{loc}},
\end{equation}
when a local exponent $\alpha_{loc} \ne \alpha$ exists, {\it i.e.} when 
anomalous roughening occurs. 
This gives rise to an anomalous temporal regime with
$\beta_{*} = \beta - \alpha_{loc}/z$.
Only when the interface is self-affine one has $\beta_* = 0$
and the Family-Vicsek scaling for the local width is recovered.
As demonstrated in Ref.\cite{lack} this may occur for growth with $\alpha$
below or above one and thus it is not necessarily
associated with super-roughening. More precisely, super-roughening corresponds
to an anomalous scaling like (\ref{anomal-scal}), but $\alpha_{loc}=1$ and
so $\beta_* = \alpha - 1/z$ in this particular case.
At true saturation, when $\l_c(t) \sim L$, the local width scales as
$$\sigma(l,t \gg L^z) \sim l^{\alpha_{loc}} L^{z \beta_*}
\sim l^{\alpha_{loc}} L^{\alpha - \alpha_{loc}},$$
that corresponds to the anomalous dependence of the local width
on the system size already observed in simulation \cite{makse}.

Next, we are going to show that for the QEW model (\ref{QEW})
the anomalous scaling behaviour of the local width
at the depinning transition
is also given by Eq.(\ref{anomal-scal}) with $\alpha_{loc}=1$ in this case.
In Figure 3 we plot a collapse of the data of Fig. 2. 
An inspection of Fig. 3   
reveals that the dynamic scaling form of the local width is consistent with 
\begin{equation}
\label{ds}
\sigma(l,t) \sim t^\beta f(l/t^{1/z}).
\end{equation}
The scaling function is given by
\begin{equation}
f(u) \sim
\left\{ \begin{array}{l}
     {\rm const} \:\:\:\: {\rm if} \:\:\:\:  u \gg 1\\
     u^{\alpha_{loc}}\:\:\:\: {\rm if} \:\:\:\:  u \ll 1,
     \end{array}
\right.
\end{equation}
where $\alpha_{loc} \simeq 0.92$, $\beta \simeq 0.81$ and $z \simeq 1.53$
are the exponents giving the best data collapse.
This scaling behavior leads to a local width 
$\sigma(l,t) \sim t^{\beta - \alpha_{loc}/z} \; l^{\alpha_{loc}}$ in 
the intermediate regime $l \ll t^{1/z} \ll L$ that corresponds to the 
scaling proposed in (\ref{anomal-scal}) with 
$\beta_* = \beta - \alpha_{loc}/z \simeq 0.21$ (to be compared with 
our previous determination $\beta_* \simeq 0.25$).  
In agreement with other authors \cite{makse,zeitak,roux}, 
we thus found $\beta \simeq 0.81$, $z \simeq 1.53$ and 
a global roughness exponent $\alpha = z \beta \simeq 1.2$
from the data collapse shown in Fig. 3.  

Similar results were obtained when the structure factor was used. 
The structure factor, which is related to the Fourier
transform of the height-height correlation function can be used to 
obtain the critical exponents
\begin{equation}
\label{spectrum}
S(k,t) = \langle \widehat{h}(k,t) \widehat{h}(-k,t) \rangle \sim 
k^{-(2\alpha + 1)} s(k^zt),
\end{equation}
where $\widehat{h}(k,t) = 
L^{-1/2} \sum_x [h(x,t) - \overline h(t)] \exp(ikx)$ and the scaling
function is 
\begin{equation}
\label{s}
s(u) \sim  
\left\{ \begin{array}{l}
     {\rm const.}  \:\:\:\: {\rm if} \:\:\:\:  u \gg 1\\
     u^{2\alpha + 1} \:\:\:\: {\rm if} \:\:\:\:  u \ll 1,
     \end{array}
\right. 
\end{equation}
whenever the interface satisfies a standard Family-Vicsek scaling. 
Also in the case of anomalous scaling due to super-roughening, Eq.(\ref{s})
gives the correct scaling of the structure factor (see [5b] for
details).  

We have determined the structure factor at the depinning transition by
performing simulations of Eq.(\ref{QEW}) in a system of total size $L=128$
and a density of pinning sites $p=1$.
For this system size, the critical force was found to be 
$F_c \simeq 0.067$. 
Figure 4 shows the data collapse for the exponents $\alpha = 1.25$ and 
$z = 1.67$ (and then $\beta = \alpha/z \simeq 0.75$) in agreement with 
the exponents that we found in Figure 3. 

\subsection{Dynamic exponent at depinning}

There is another point that merits some discussion related to the complete
scaling of the QEW model. From Figs. 2 and 3 we have obtained a value for the
dynamic exponent $z \approx 1.5-1.6$ in agreement with 
several previous simulations \cite{makse,zeitak}
of the QEW equation (and related models) in which an
exponent $z \simeq 1.5$ was reported. However, in these numerical works
$z$ was always calculated from the scaling relation
$z=\alpha/\beta$, as we have done in this paper.
In principle, in order to correctly determine the dynamic
exponent $z$ one should be able to use
its definition in $\l_c(t) \sim t^{1/z}$, where
the correlation length $\l_c(t)$ can be obtained
from an appropriate correlation function.
Let us consider the height correlation function
$\Gamma(l,t) = \left \{ \langle \bigl(h(x+ l,t) -
\langle h\rangle \bigr) \bigl(h(x,t) - 
\langle h \rangle\bigr)\rangle \right \},
$
which relates the heights at two positions $x$ and $x+l$.
This correlation function becomes zero at distances $l$ larger than
the correlation length, thus $\l_c(t)$ can be determined from
\begin{equation}
\label{corre-leng}
\l_c(t) = {\int l \Gamma(l,t) dl \over \int \Gamma(l,t) dl}\sim t^{1/z}.
\end{equation}
Note that Eq.(\ref{corre-leng}) should be the correct way of obtaining $z$
while we are interested in checking the validity of the dynamic
scaling ansatz. Direct determination
of the dynamic exponent from Eq.(\ref{corre-leng}) was
reported in Ref.\cite{europhys,jost}, where $z' \simeq 2$ for a wide range of
driving forces even at the transition 
(here the prime denotes a direct measure using Eq.(\ref{corre-leng})). 
Surprisingly, the value of the exponent $z' \simeq 2$  
obtained in this way differs from the one that we have
measured by an {\em indirect} method $z \simeq 1.5$ (and consecuently 
the corresponding global roughness exponents $\alpha \simeq 1.2$ and 
$\alpha' \simeq 1.5$ are different as well).     
A value of the dynamic exponent $z' \simeq 2$ 
and the time exponent $\beta \simeq 0.75$ leads to
$\beta_* = \beta - 1/z' \simeq 0.25$, in agreement with our previous 
estimation for the anomalous time exponent.

In order to try to explain 
this puzzeling situation we have considered the effect
of the characteristic size of the pinned regions on the 
interface local scaling.     
Besides the super-rough character of interfaces generated by the QEW
equation and the corresponding anomalous scaling of the local width just
discussed above,
there is also another important effect that conspires to
complicate even more the QEW problem.
As we have already mentioned, the quenched disorder in Eq.(\ref{QEW})
can slow down the motion of the interface in large regions of length $\xi $
in which the interface temporarily remains pinned.
In an infinite system, the size
of the pinned regions, $\xi$, diverges as $\xi \sim (F-F_{c})^{-\nu} $
when approaching the depinning threshold $F_c$ from above
and the exponent $\nu \simeq 1.35$ \cite{makse}
(or $\nu \simeq 4/3$ in Ref.\cite{nolle})
can be measured in simulation
from a direct observation of these pinned regions in the interface.
For a system of finite size $L$
the depinning transition occurs at a value of the driving force
$F=F_{d}(L)$ for which $(F_{d}-F_{c})^{-\nu} \sim L $. As
a consequence, in a finite system
the depinning force $F_{d}(L)$ depends on the system
size, and close to the depinning transition both lengths $\xi$ and
$L$ become relevant in the dynamics.
Thus, in any proper numerical determination of
the global roughness exponent at the depinning threshold one must
have in mind two important facts. First, to obtain the global
exponent $\alpha'$ (where $\alpha' = z'\beta \approx 1.5$) 
one has to perform simulations in systems with different
sizes $L$ and, after saturation for $t \gg L^z$, one can fit the power law
$\sigma (L, t \gg L^z) \sim L^{\alpha'}$.
Secondly, since the critical force $F_d(L)$ is different for a different
size of the system,
in order to keep the system at the depinning transition
one must vary the driving force $F$ in such a way that $\xi \sim L$.
These requirements lead to a dependence of the
saturated global width on the relevant lengths $\xi$
and $L$ at the depinning transition as
\begin{equation}
\sigma (L, t \gg L^z) \sim L ^{\alpha'} ({\xi \over L })^ \theta
\sim L^{\alpha' - \theta} (F_d(L) - F_c)^{-\nu \theta},$$
\label{anomal-global}
\end{equation}
where $\theta$ is an exponent that characterizes the
dependence of the global width on the correlation length $\xi$.
It is worth remarking here that only when $\xi \sim L$ the system is
certainly being maintained at
the depinning transition and the exponent $\alpha'$ can be measured.
On the contrary,
numerical simulations done in systems with different sizes, but in which
the driving force $F$ is not adjusted to the corresponding $F_d(L)$, always
yield a different roughness exponent $\alpha = \alpha' - \theta$.
The exponent $\alpha$ would appear for any $F$ somewhat larger than $F_d$
and gives the global roughness in the {\em moving} phase.

In Ref.\cite{makse} an
anomalous dependence of the prefactor of the width on the driving force
as $(F - F_c)^{-\phi}$ was already found, with $\phi \simeq 0.44$.
We believe that the dependence of the saturated width on the driving force
close to the depinning transition indicates
that the system is not just at the transition. Taking into account the
measured values $\nu \simeq 1.35$ \cite{martys,makse} 
and $\phi \simeq 0.44$ \cite{makse}
we have $\theta = \phi/\nu \simeq 0.33$.
So, we obtain a global roughness exponent
$\alpha= \alpha' - \theta \simeq 1.17$ in agreement with our numerical 
determination 
($\alpha \simeq 1.2$, see Fig. 3 and Fig. 4) as well as with previous
simulations \cite{roux,makse,jensen,leschhorn,galluccio}.

\section{discussion}
In summary, we have found
that the complete scaling behaviour of
the QEW equation is not trivial. At the depinning
transition the interface is super-rough and this leads to
the violation of the usual scaling ansatz. In particular, the local width
exhibits a new time regime with an exponent $\beta_* \simeq 0.21$.
We have shown that
this is associated with the existence of two different roughness exponents
describing the scaling of the local and global fluctuations.
A new dynamic scaling, which
was already used in other growth processes, has been
successfully applied to understand the scaling behaviour of the
local width.
 
Accordingly to this scaling, Eq.(\ref{anomal-scal}), 
the measured value of the local 
roughness, close to unity $\alpha_{loc} \simeq 0.92$, suggests that
the anomalous scaling of the local width is due to super-roughening,
in such a way that a global roughness exponent larger than one must exist.
A direct determination of the time exponent gives $\beta \simeq 0.75-0.81$
in agreement with most of the previous works. On the contrary, a direct
determination of the global roughness exponent from the scaling of the 
saturated global width with the system size can be controversial since 
variation of the total size $L$ leads to a change in the depinning 
threshold $F_c$. The effect of the characteristic 
length of the pinned sites regions
is relevant close to the depinning transition. It seems then that it is 
possible to obtain two different values for the global roughness exponent 
depending on the procedure used.
The most usual reported value of the 
global roughness exponent $\alpha \simeq 1.2$ (and $z \simeq 1.5$) is 
obtained by us from collapses of the local width as well as from the 
behaviour of the structure factor. However, a direct measure of the 
correlation lenght reflects that there is a dynamic exponent $z' \simeq 2$
\cite{europhys,jost} (and consequently $\alpha' \simeq 1.5$). 
Moreover, it remains unclear
for us which of the mentioned methods (and the corresponding set of 
critical exponents) is the proper to describe completely the scaling
behaviour of the interface at depinning. 
We believe that it is a very unsatisfactory situation the fact that, on the one
hand, the depinning threshold changes with the system size and, on the other
hand, one of the exponents of interest, the global roughness exponent,
has to be measured by changing the system size. It seems difficult to 
define a correct roughness exponent at critical depinning as the exponent 
$\alpha$ in the power law $\sigma(L,t \to \infty) \sim L^\alpha$ while a 
variation of $L$ yields a shift in the critical point.

From an experimental point of view
the local roughness exponent is very important
since it is the exponent accessible in experiments, in which the
size of the sample remains fixed. We suggest that the anomalous scaling
(\ref{anomal-scal}) might be useful for clarifying whether the QEW equation
certainly describes the motion of interfaces in disordered media, since in
many experiments it may be possible to study the time evolution
of the local width and to determine the existence of an
anomalous exponent $\beta_*$.

\acknowledgments
We thank Rodolfo Cuerno for discussions and encouragement . This work has been 
supported by DGICyT of the Spanish Government under Project 
No. PB93-0054-C02-02. J.\ M.\ L.\ acknowledges finantial support from 
Postdoctoral Program of Universidad de Cantabria to work at Instituto de 
Fisica de Cantabria, where much of this work has been made.

\begin{figure}
\centerline{
\epsfxsize=5cm
\epsfbox{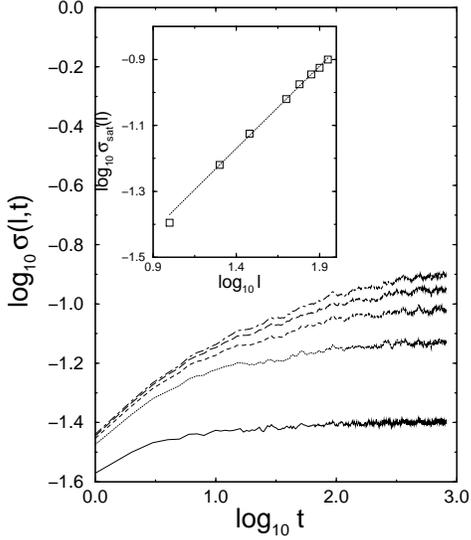}}
\caption{Time evolution of the local width on different
length scales $l$ (window size) for a driving force $F=0.4$, which is
far from the depinning transition.
Local width saturates at longer times for
larger length scales as corresponds to a standard scaling.
The values of the width in saturation are taken and ploted
vs. scale $l$ in a log-log plot (see inset).
The slope of the line (inset) that fits the numerical data is 0.5
and gives a local roughness $\alpha_{loc} \simeq 0.5$
}
\end{figure}

\begin{figure}
\centerline{
\epsfxsize=5cm
\epsfbox{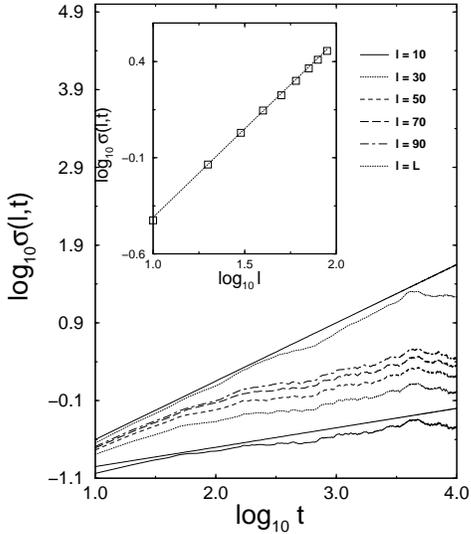}}
\caption{Time evolution of the 
local width at the depinning transition,
$F=F_c \simeq 0.13$. The global width, $l=L$, is also displayed.
The continuous straight lines have slopes 3/4 and 1/4 and 
are shown to guide the eye.
A simple comparison with Fig.1 reveals the anomalous scaling behaviour
for this case.
The local width does not saturate, but crosses over to an
anomalous time regime, $t^{\beta_*}$,     
where $\beta_* \simeq 0.25$
and saturation occurs at the same time for the whole system.
Inset shows the width in true saturation (flat part in the plots)
vs. the scale $l$ in a log-log plot. The line (inset) fits the data
and the slope corresponds to $\alpha_{loc} \simeq 0.92$.
}
\end{figure}

\begin{figure}
\centerline{
\epsfxsize=5cm
\epsfbox{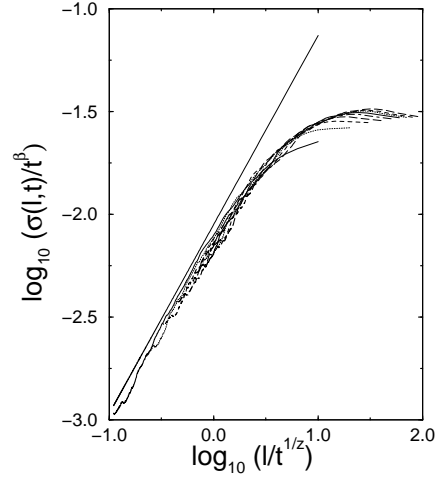}}
\caption{Data collapse, according to (9), of the results displayed in Fig. 2.
The exponents $\beta = 0.81$ and $z = 1.53$ were used. The slope of the 
straight line gives the local roughness exponent $\alpha_{loc} = 0.92$.}
\end{figure}

\begin{figure}
\centerline{
\epsfxsize=5cm
\epsfbox{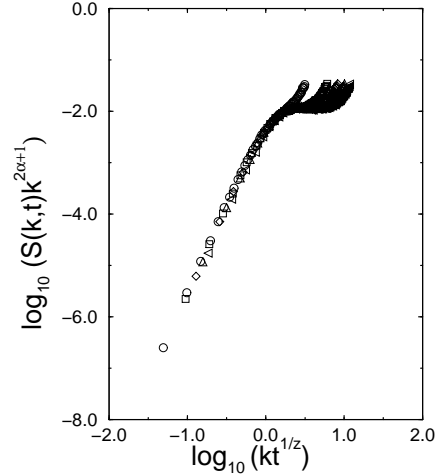}}
\caption{Structure factor data collapse for the QEW equation at the 
depinning transition in a system of size $L=128$. Data shown a good collapse
for $z=1.67$ and $\alpha = 1.25$. The deviation from the scaling at large 
abcissa values is due to discrete lattice effects.}
\end{figure}

\end{multicols}


\begin{references}

\bibitem{general}
 F.Family T.Vicsek eds.,
 {\em Dynamics of Fractal Surfaces}
 (World Scientific, 1991, Singapore);J. Krug and H. Spohn, in {\em Solids Far
 From Equilibrium: Growth, Morphology and Defects}, edited by C. Godr\'eche
 (Cambridge University Press, Cambridge, 1991).

\bibitem{review}
T. Halpin--Healey and Y.-C. Zhang, Phys. Rep. {\bf 254}, 215 (1995);
A.-L. Barabasi and H.E. Stanley, {\em Fractal concepts in surface growth},
(Cambridge University Press, Cambridge, 1995).

\bibitem{ew}
S.F. Edwards and D.R. Wilkinson,
Proc. R. Soc. Lond. {\bf A381}, 17 (1982).

\bibitem{kpz}
M. Kardar, G. Parisi and Y.-C. Zhang,
Phys. Rev. Lett.  {\bf 56}, 889  (1986).

\bibitem{lack} J.M. L\'opez and M.A. Rodr\'\i guez, Phys. Rev. E, 
{\bf 54}, R2189 (1996); J.M. L\'opez, M.A. Rodr\'\i guez and 
R. Cuerno, preprint cond-matt/9703024 (to be published).

\bibitem{das-sarma} S. Das Sarma, C.J. Lanczycki, R. Kotlyar,
and S.V. Ghaisas, Phys. Rev. E {\bf 53}, 359 (1996).

\bibitem{exper}
M.A. Rubio, C.A. Edwards, A. Dougherty and J.P. Gollub,
Phys. Rev. Lett.  {\bf 63}, 1685 (1989);
V.K. Horvath, F. Family and T. Vicsek,
J. Phys. A: Math. Gen.  {\bf 24}, L25  (1991);
V.K. Horvath, F. Family and T. Vicsek,
Phys. Rev. Lett.  {\bf 65}, 1388 (1990);
S. He, G.L.M.K.S. Kahanda and Po-zen Wong,
Phys. Rev. Lett.  {\bf 69}, 3731 (1992).

\bibitem{RG}
T. Nattermann, S. Stepanow, L.-H. Tang and H. Lesch\-horn,
J. Phys. II France  {\bf 2}, 1483 (1992);
O. Narayan and D.S. Fisher,
Phys. Rev. B {\bf 48}, 7030 (1993).

\bibitem{anomalo}
J.M. L\'opez, Phys. Rev. E {\bf 52}, R1296 (1995).

\bibitem{europhys}
J.M. L\'{o}pez, M.A.
Rodr\'{\i}guez, A. Hern\'{a}ndez-Machado, and
A. D\'{\i}az-Guilera, Europhys. Lett., {\bf 29}, 197, (1995)

\bibitem{roux}
S. Roux and A. Hansen, J. Phys. I (France) {\bf 4}, 515 (1994).

\bibitem{jensen}
H.J. Jensen, J. Phys. A {\bf 28}, 1861 (1995).

\bibitem{makse} H.A. Makse and L.A.N. Amaral,
Europhys. Lett. {\bf 31}, 379 (1995); L.A.N. Amaral, A.-L. Barab\'asi,
H.A. Makse and H.E. Stanley, Phys. Rev. E {\bf 52}, 4087 (1995).

\bibitem{nolle}
C.S. Nolle, B. Koiller, N. Martys and M.O. Robbins,
Phys. Rev. Lett. {\bf 71}, 2074 (1993).

\bibitem{kessler}
D.A. Kessler, H. Levine and Y. Tu,
Phys. Rev. A  {\bf 43}, 4551 (1991).

\bibitem{martys}
N. Martys, M. Cieplak and M.O. Robbins,
Phys. Rev. Lett.  {\bf 66}, 1058 (1991).

\bibitem{galluccio}
S. Galluccio and Y.-C. Zhang, Phys. Rev. E {\bf 51}, 1686 (1995).

\bibitem{csahok}
Z. Csah\'ok, K. Honda, E. Somofai, M. Vicsek, and T. Vicsek,
Physica A {\bf 200}, 136 (1993).

\bibitem{leschhorn}
H. Leschhorn and L.-H. Tang,
Phys. Rev. Lett. {\bf 70}, 2973 (1993)

\bibitem{zeitak}
Z. Olami, I. Procaccia, and R. Zeitak, Phys. Rev. E {\bf 52}, 3402 (1995).

\bibitem{jost}
M. Jost and K.D. Usadel, 
Phys. Rev. B {\bf 54}, 9314 (1996);
M. Jost and K.D. Usadel, 
Physica A (to appear).

\end{references}
\end{document}